\documentclass[twocolumn,showpacs,preprintnumbers,amsmath,amssymb]{revtex4}
\usepackage{graphicx}
\usepackage{dcolumn}
\usepackage{bm}
\usepackage{amsmath}

\begin{document}

\title{Theoretical rate of dissociative recombination of HCO$^{+}$ and DCO$^{+}$ ions}
\author{Nicolas Douguet, Viatcheslav Kokoouline}
\affiliation{Department of Physics, University of Central Florida, Orlando, Florida 32816}
\author{Chris H. Greene}
\affiliation{Department of Physics and JILA, University of Colorado, Boulder, Colorado 80309-0440}


\begin{abstract} 
The article presents an improved theoretical description of dissociative recombination of HCO$^+$ and DCO$^+$ ions with a low energy electron. In the previous theoretical study (Phys. Rev. A {\bf 74}, 032707) on HCO$^+$, the vibrational motion along the CO coordinate was neglected. Here, all vibrational degrees of freedom, including the CO stretch coordinate, are taken into account. The theoretical dissociative recombination cross-section obtained  is similar to the previous theoretical result at low collision energies ($<$0.1 eV) but somewhat larger at higher ($>$0.1 eV) energies. Therefore, the present study suggests that motion along the CO coordinate does not play a significant role in the process at low collision energies. The theoretical cross-section is still approximately 2-3 times lower than the data from a recent merged-beam experiment.
\end{abstract}


\maketitle

The HCO$^+$ ion has been known for more than a century and was the first ion discovered in interstellar space via microwave spectroscopy \cite{Buhl}. It was classified as unidentified element at the time of discovery. Later Klemperer \cite{Klemperer} suggested the linear HCO$^+$ ion as a candidate, which was later confirmed by experiments (see, for example Ref. \cite{Woods}). These last two decades, HCO$^+$ and other small polyatomic ions have extensively been studied theoretically: their electronic structure, potential energy surfaces and equilibrium geometry have been systematically investigated as, for example, in  Refs. \cite{Yamaguchi,Palmieri}. The spectroscopy of neutral HCO in the energy range relevant to these DR studies has been extensively mapped out by Grant and coworkers \cite{GrantV}. Dissociative recombination (DR) of molecular ions like HCO$^+$ plays an important role in chemistry of interstellar clouds and therefore allows astronomers to probe remotely various characteristics of these clouds. In space, the HCO$^+$ ion can be formed by several possible associations such as H$_2$ + CO$^+$ or CH + O and also H$_3^+$ + CO \cite{rowe92,Mcgregor} and destroyed by DR. Different types of laboratory experiments have been performed in order to study DR of HCO$^+$: afterglow plasma, merged-beam and storage ring experiments \cite{rowe92,Gougousi,adams84,lepadellec97,amano90,poterya05}. From the laboratory experiments it is now known that DR in HCO$^+$ proceeds mainly into the H+CO channel:  HCO$^+$ +e$^-$ $\rightarrow$ H + CO. On the other hand, at present, there is no consensus among different experimental measurements of the actual DR rate coefficient: as they differ by up to a factor of ten \cite{lepadellec97,amano90}.

The theory of DR in diatomic ions has been reasonably well developed in recent decades. For triatomic ions, only recently has theory been able to provide meaningful results for the simplest triatomic ion, H$_3^+$  \cite{orel93,kokoouline01,kokoouline03a}. The theoretical description of DR in triatomic molecular ions is a difficult problem in part because several different (electronic and vibrational) degrees of freedom have to be taken into account. Several approximations have also been made in a recent theoretical study of DR in HCO$^+$  \cite{Ivan}. The obtained DR cross-section was about factor 2.5 smaller than the lowest experimental cross-section \cite{lepadellec97}. One of the possible reasons why the theoretical cross-section was smaller in Ref. \cite{Ivan} than the experimental one is the approximation of the frozen CO coordinate fixed at its equilibrium value. Although the main dissociation pathway does not involve this coordinate, it was argued \cite{Ivan} that the CO vibration could increase the probability to capture the electron and increase the overall DR cross-section. In the present study, we improve the previously developed DR treatment in HCO$^+$ and investigate the explicit role of the CO vibration. 

The theoretical treatment presented here resembles in many respects the approach applied previously to the H$_3^+$ \cite{kokoouline01} and HCO$^+$ \cite{Ivan} target ions. Below, we describe the new elements of the theoretical approach. We represent the Hamiltonian of the ion+electron system as $H=H_{ion}+H_{el}$, where $H_{ion}$ is the ionic Hamiltonian and $H_{el}$ describes the electron-ion interaction. Consider first the ionic Hamiltonian $H_{ion}$ written in the center-of-mass reference frame. We use Jacobi coordinates to represent all vibrational degrees of freedom: Introducing {\it G} as the center of mass of C-O, the set of internuclear coordinates is represented by the quartet $\mathcal{Q}$=$\{ R_{\rm{CO}},R_{G\rm{H}},\theta,\varphi \}$. Here $R_{\rm{CO}}$ and $R_{G\rm{H}}$ represent respectively the distances C-O and {\it G}-H, $\theta$ is the bending angle between $\overrightarrow{\rm{OC}}$ and $ \overrightarrow{G\rm{H}}$, $\varphi$ is the azimuthal orientation of the bending. Here, we consider $R_{G\rm{H}}$ as the adiabatic coordinate representing the dissociation path. In the previous study \cite{Ivan}, the inter-nuclear distance $R_{\rm{CO}}$ was fixed at its equilibrium value ($R_{\rm{CO}}$=2.088 a.u) and the  $R_{\rm{CH}}$ coordinate was treated as the dissociative coordinate.  Note that even though we use an adiabatic representation, our inclusion of nonadiabatic coupling effects means that we are not utilizing an adiabatic approximation since the adiabatic eigenstates form a complete basis set.

\begin{figure}[ht]
\includegraphics[width=8cm]{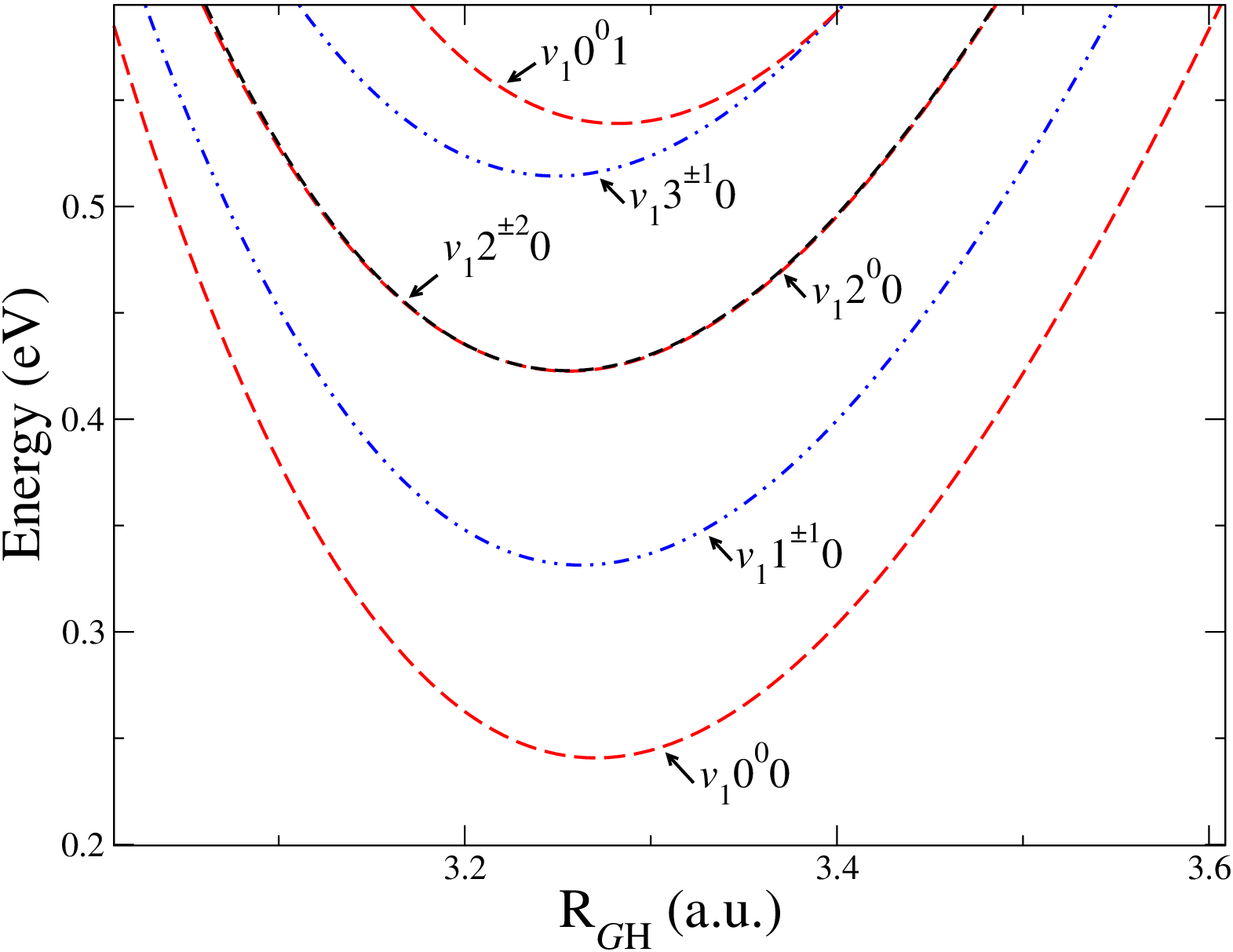}
\caption{(Color online) Adiabatic curves $U_{m_{\varphi},l}(R_{G\rm{H}})$ versus the adiabatic coordinate $R_{G\rm{H}}$. The curves are labeled with quantum numbers $\{v_1v_2^{m_\varphi}v_3\}$ of normal modes of HCO$^+$. The $v_1$ quantum in our model corresponds approximately to motion along $R_{G\rm{H}}$ and therefore is not defined for the adiabatic curves because $R_{G\rm{H}}$ is not quantized. The curves $\{v_12^{\pm2}0\}$ and $\{v_12^00\}$ are almost degenerate.}
\label{fig:ionic_potential}
\end{figure}

The vibrational Hamiltonian in the Jacobi coordinates is:
\begin{eqnarray}
\label{eq:hamiltonian}
H_{ion}={-\frac{\hbar^2}{2\mu_{\rm{CO}}}\frac{\partial^{2}}{\partial R_{\rm{CO}}^{2}}} -{\frac{\hbar^2}{2\mu_{\rm{H-CO}}}\frac{\partial^{2}}{\partial R_{\rm{H-CO}}^{2}}}+\nonumber\\
+{\frac{\hat L^{2}(\theta,\varphi)}{2\mu_{\rm{H-CO}} R_{\rm{H-CO}}^{2}}}+{V(R_{\rm{CO}},R_{G\rm{H}},\theta)}\,,
\end{eqnarray}
where $\mu_{\rm{CO}}$ and $\mu_{\rm{H-CO}}$ are respectively the reduced masses of the C--O and H--CO pairs; $\hat L^{2}(\theta,\varphi)$ is the familiar operator of angular momentum corresponding to relative rotation of H and the CO axis. Representing the ion by the above Hamiltonian depending on $\{ R_{\rm{CO}},R_{G\rm{H}},\theta,\varphi \}$ only, we neglected by rotational motion of the CO bond in space, but included relative rotation of H and CO. This approximation is justified by a large CO/H mass ratio. As a result of the approximation, the projection $m_{\varphi}$ of the angular momentum $\hat L$ on the CO axis is conserved. We solve the Schr\"odinger equation with the Hamiltonian (\ref{eq:hamiltonian}) keeping the $R_{G\rm{H}}$ coordinate fixed; this determines vibrational wave functions $\Phi_{ m_{\varphi},l}(R_{G\rm{H}};R_{CO},\theta,\varphi)$ and corresponding adiabatic energies $U_{m_{\varphi},l}(R_{G\rm{H}})$ that depend parametrically on $R_{G\rm{H}}$. Several of these curves are shown in Fig. \ref{fig:ionic_potential}. The lowest adiabatic curves can approximately be characterized by quantum numbers \{$v_{1},v_{2}^{m_\varphi},v_{3}$\} of the four normal modes of the HCO$^+$ ion. 

A reasonable measure of accuracy of the adiabatic approximation is provided by comparing the energy splitting between our adiabatic potentials $U_{m_{\varphi},l}(R_{G\rm{H}})$ with exact calculations of the corresponding vibrational splittings \cite{Palmieri} (Table \ref{tab:vibr_ene}). The obtained values are practically the same as in the previous study \cite{Ivan} also shown in the table, since a similar (but not identical) adiabatic approximation was used in that study.

\begin{table}[tbp]
\begin{tabular}{|p{1.5cm}|p{2cm}|p{2cm}|p{2.cm}|}
\hline
 $\{v_1v_2^{l},v_3\}$ & Present calculation & Previous calculation\cite{Ivan}& Puzzarini {\it et al.}\cite{Palmieri} \\ 
\hline\hline
$10^00$ &343& 363 & 383.1\\
$01^10$ &91 & 92 &  103.0 \\
$02^00$ &182 & 181 &  203.5\\
$03^10$ &275  & 273 &  304.9 \\
$04^00$ &369  & 362 &  403.8 \\
$00^01$ &298  & no value &  270.6 \\
\hline
\end{tabular}
\caption{Comparison of vibrational energies obtained in the adiabatic approximation with the exact calculation from Ref. \cite{Palmieri}. The result of the previous study \cite{Ivan}, where a different adiabatic approximation was used is also shown. In that study CO was not quantized and thus, $00^01$ was not calculated. The overall error is about $12 \%$, which translates into about  25 $\%$ for vibrational wave functions. The energies are  given in meV.}
\label{tab:vibr_ene}
\end{table}

The  structure of the electronic part $H_{el}$ of the total Hamiltonian $H$ is the same as in the previous study \cite{Ivan}: It includes $ns\sigma $, $np\pi ^{-1}$, $np\sigma $, $np\pi ^{+1}$, and $nd\sigma $ electronic states only. In the basis of the five electronic states, the Hamiltonian has the following block-diagonal form for each principal quantum number $n$: 
\begin{equation}
H_{int}(\mathcal{Q})=\left( 
\begin{array}{ccccc}
E_{s\sigma } &0 & 0 & 0 &0\\ 
0&E_{p\pi }  & \delta e^{i\varphi } & \gamma e^{2i\varphi } &0\\ 
0& \delta e^{-i\varphi } & E_{p\sigma }& \delta e^{i\varphi } &0\\ 
0&\gamma e^{-2i\varphi } & \delta e^{-i\varphi } & E_{p\pi }&0\\ 
0&0 & 0 & 0&E_{d\sigma }
\end{array}
\right) \,,  
\label{eq:Hint}
\end{equation}
where $E_{s\sigma }$, $E_{p\sigma }$, $E_{p\pi }$, and $E_{d\sigma }$ are the energies of the corresponding electronic states at the linear ionic configuration; $\delta $ and $\gamma $ are the real non-Born-Oppenheimer coupling parameters. The couplings $\delta $ and $\gamma $ depend on $R_{G\rm{H}},\ R_{\rm{CO}}$, and $\theta $ and responsible for the Renner-Teller interaction. They are zero for linear geometry of the ion. The parameters in the above Hamiltonian are obtained from {\it ab initio} calculations of Ref. \cite{larson05} as discussed in Ref. \cite{Ivan}. In the present method, the electron-ion interaction Hamiltonian  $H_{int}(\mathcal{Q})$ is now used to construct the $5\times 5$ reaction matrix $K_{i,i'}(\mathcal{Q})$ written in the same basis of electronic states as $H_{int}(\mathcal{Q})$. 

Once the adiabatic states $\Phi_{ m_{\varphi},l}(R_{G\rm{H}};R_{\rm{CO}},\theta,\varphi)$, energies $U_{m_{\varphi},l}(R_{G\rm{H}})$, and the reaction matrix $K_{i,i'}(\mathcal{Q})$  are obtained, we take $R_{G\rm{H}}$ as the adiabatic coordinate and apply the quantum-defect approach that has been already used in a number of DR studies of triatomic and diatomic ions \cite{kokoouline01,moshbach05}. We construct the reaction matrix $\mathcal{K}_{j,j^{\prime }}(R_{G\rm{H}})$
\begin{equation}
\mathcal{K}_{\{m_\varphi,l,i\},\{m_\varphi,l,i\}^{\prime }}(R_{G\rm{H}})=\langle \Phi _{m_\varphi,l}|K_{i,i'}(\mathcal{Q})|\Phi _{m' _\varphi,l'}\rangle\,,
\end{equation}
where the integral is taken over the three internuclear coordinates, $R_{\rm{CO}}$, $\varphi$, and $\theta$.
The reaction matrix $\mathcal{K}_{j,j^{\prime }}$ thus obtained has many channels and parametrically depends on $R_{G\rm{H}}$. For each $R_{G\rm{H}}$ value, we then obtain a number of resonances with energies $U_a(R_{G\rm{H}})$ and widths $\Gamma_a(R_{G\rm{H}})$. The resonances correspond to the autoionizing electronic states of the neutral molecule at frozen $R_{G\rm{H}}$. The fixed-$R_{G\rm{H}}$ width of the resonances is the reciprocal of the fixed-$R_{G\rm{H}}$ resonance autoionization lifetime, which of course is not an experimentally-observable resonance width since $R_{G\rm{H}}$ has not been quantized.

The energies and widths of the resonances are then used to calculate the cross-section for electron capture by the ion. Depending on whether or not a particular neutral potential curve  $U_a(R_{G\rm{H}})$ is energetically open for direct dissociation, two different formulas are appropriate to use for the cross-section calculation. For the neutral states energetically open for direct dissociation, we have
\begin{equation}
\label{eq:DA}
 \sigma=  \frac{2\pi^2}{k_o^2}
         \sum_a\frac{\Gamma_{a}(R_{G\rm{H}})}{|U'_a(R_{G\rm{H}})|}|\chi^+_o(R_{G\rm{H}})|^2\,.
\end{equation}
However, the following equation should be used 
\begin{equation}
\label{eq:sum_over_channels}
\sigma=\frac{2\pi^2}{k_o^2}\sum_c|\langle\chi^{\rm res}(R_{G\rm{H}}) | \sqrt{\Gamma_{a}(R_{G\rm{H}})}|\chi^+_o(R_{G\rm{H}})\rangle|^2 n_c^3\,
\end{equation}
for the  $U_a(R_{G\rm{H}})$ curves that are energetically closed to direct dissociation \cite{Ivan}.
In the above equations, $k_o$ is the asymptotic wave number of the incident electron, which  depends on the initial state $o$ of the target molecular ion; $\chi^+_o(R_{G\rm{H}})$ is the initial vibrational wave function of the ion; $\chi^{\rm res}(R_{G\rm{H}})$ is the quantized radial wave function of the  $U_a(R_{G\rm{H}})$ curve. The sum in the first equation includes all neutral states open for direct dissociation. The sum in the above equation is over all closed channels $c$ that produce potential curves $U_a(R_{G\rm{H}})$ closed to direct dissociation \cite{Ivan}.

The total DR cross-section for HCO$^+$ is mainly determined by the second sum because the electron is most likely captured  into one of the lowest closed channels that cannot dissociate directly. As mentioned previously, the formulas above describe the cross-section for capture of the electron. It is equal to the DR cross-section only if the probability of subsequent autoionization is negligible compared to the dissociation probability, after the electron has been captured by the ion.

The projection $M=m_\varphi+\lambda$ of the total angular momentum on the CO molecular axis, where $\lambda$ is the projection of the electronic angular momentum on the CO axis, is a conserved quantity in our model. Therefore, the resonances and the cross-section are calculated separately for each value of $|M|$. Since $\lambda$ can only be 0 or $\pm$1 in our model ($\sigma$ and $\pi$ states) and the initial vibrational state of the ion has $m_{\varphi}=0$, the possible values of $|M|$ are 0 and 1. The total cross-section for electron capture by the ion in the ground vibrational level is given \cite{Ivan} by $\langle\sigma^{total}\rangle=\langle\sigma^{M=0}\rangle + 2\langle\sigma^{M=1}\rangle$.

\begin{figure}[h]
\includegraphics[width=8cm]{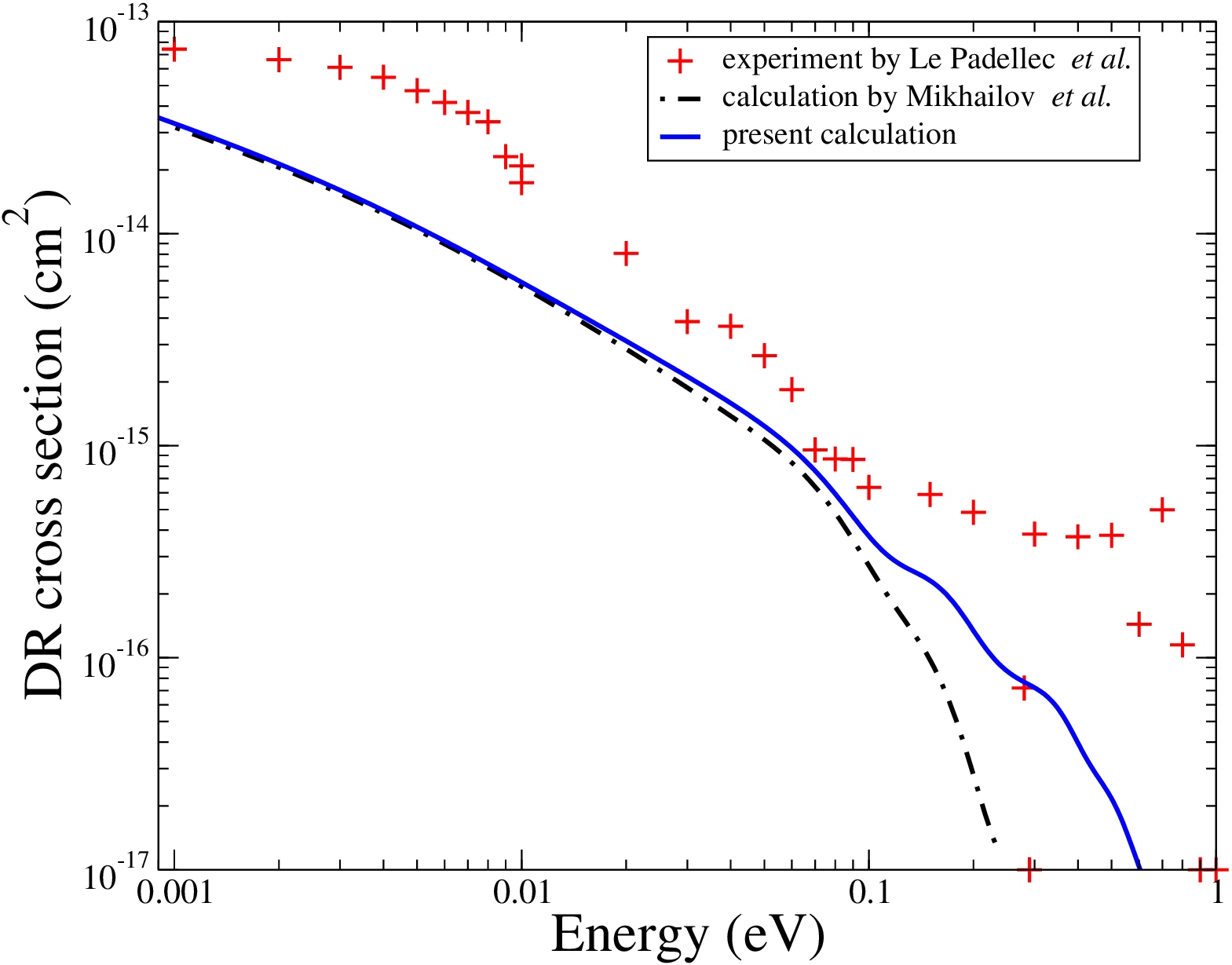}
\caption{(Color online) The figure shows the calculated DR cross-section for HCO$^+$ (solid line) as a function of the incident electron energy. The experimental \protect\cite{lepadellec97} (cross symbols) and previous theoretical \cite{Ivan} (dashed line) cross-sections are also shown for comparison. The theoretical curves include a convolution over the experimental electron energy distribution according to the procedure described in Ref. \cite{moshbach05} with $\Delta E_{\perp}=\Delta E_{||}=3$ meV. }
\label{fig:merged-beam}
\end{figure}
\begin{figure}
\includegraphics[width=8cm]{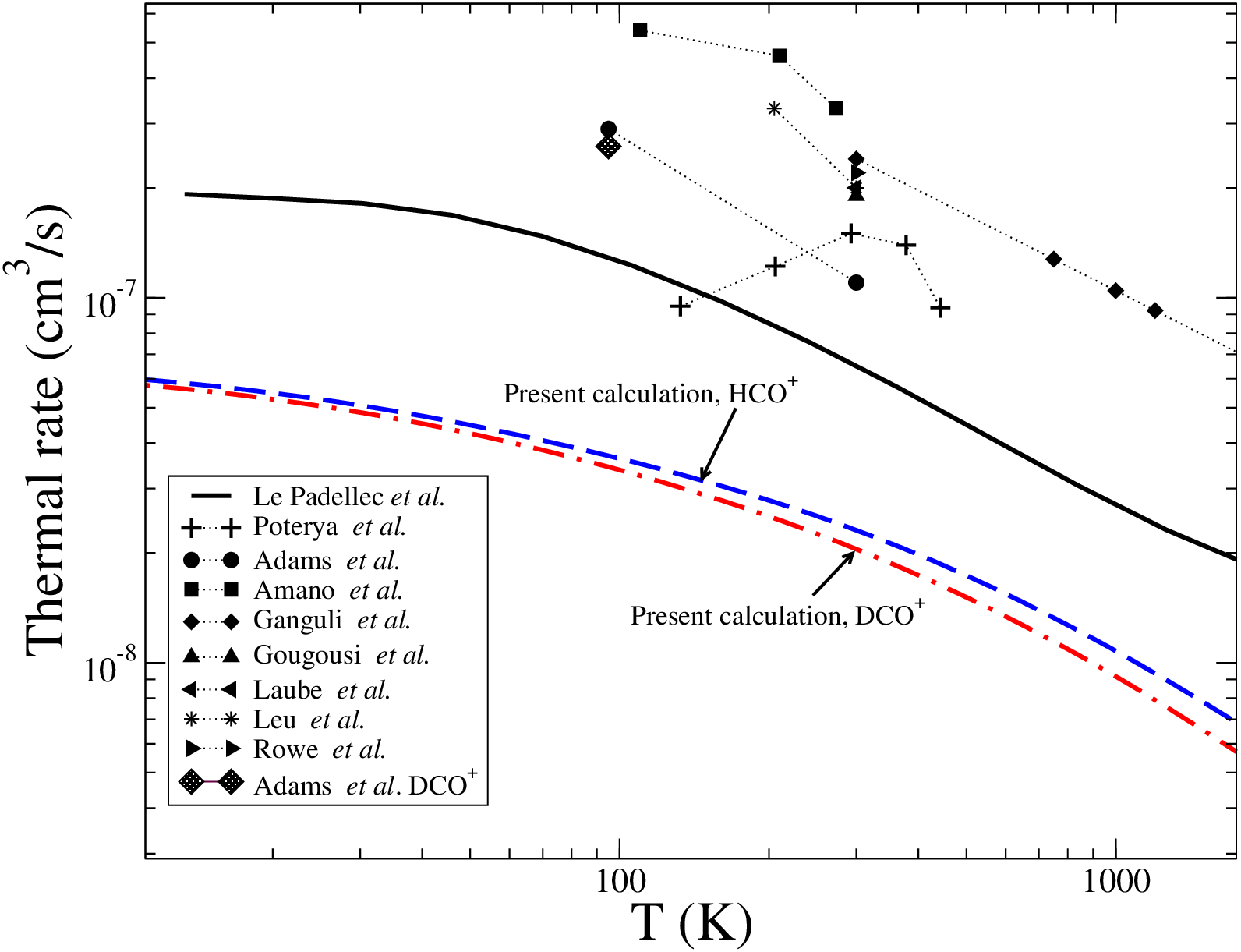}
\caption{(Color online) Theoretical (dashed lines) and experimental DR thermal rates for HCO$^+$ and DCO$^+$. The only available experimental data point for DCO$^+$ is shown as a diamond symbol. The other symbols and the solid line represent data from experiments with HCO$^+$. }
\label{fig:thermal-rate}
\end{figure}

 
Figures \ref{fig:merged-beam} and \ref{fig:thermal-rate} summarize the results of the present calculation. Fig. \ref{fig:merged-beam} compares the present results with the experimental data from a merged-beam experiment \cite{lepadellec97} and with the previous theoretical study \cite{Ivan}. The theoretical results are almost identical (about 10\% different) for electron energies below 0.1 eV.  However they differ significantly at higher energies, where the present calculation gives a higher cross-section. Both curves are smaller than the experimental data by a factor of 2-3. Therefore, the approximation of the frozen C-O bond employed in Ref. \cite{Ivan} is apparently justified for low electron energies but appears to deteriorate at higher energies. This result can be rationalized as follows. For small electron energies, the CO vibration plays a negligible role because only a few resonances are associated with excited CO vibrational modes. In addition, normally, widths of these resonances are relatively small due to small relevant coupling in the corresponding reaction matrix elements: The largest coupling elements in the matrix are associated with the Renner-Teller coupling, which is active when $m_\varphi$ is changed. However, when the total energy of the system becomes close to (but below) the first CO-excited level $\{00^01\}$ of the ion (0.3 eV above the ground vibrational level), the Rydberg series of resonances associated with the $\{00^01\}$ level becomes more dense and, more importantly, they become mixed with the Rydberg series of the resonances associated with  $\{03^10\}$.  The latter are coupled relatively strongly to the ground vibrational level $\{00^00\}$ of the ion by the Renner-Teller coupling.

Figure \ref{fig:thermal-rate} shows the thermal rate coefficients obtained in the present study for HCO$^+$ and DCO$^+$ and compares them with available experimental data. Somewhat analogous to Fig. \ref{fig:merged-beam}, the theoretical DR rate coefficient for HCO$^+$ is smaller than the rate coefficient obtained from the merged-beam experiment \cite{lepadellec97} by about a factor of 3. The majority of the other experimental thermal rate coefficients shown in the figure are obtained in plasma experiments. These rates are significantly higher than the merged-beam experimental data \cite{lepadellec97}.

In the previous treatment \cite{Ivan}, the DR rate coefficient obtained for DCO$^+$ was approximately 30\% smaller than for HCO$^+$. In the present study, the  DCO$^+$ coefficient is smaller than the one in HCO$^+$ by only 10\%. The only experimental data available for DCO$^+$ is the rate coefficient $2.6\times10^{-7}$cm$^{3}/s$ obtained at T=95~K  \cite{adams84}. The same study provides the rate coefficient for HCO$^+$, which is by 10\% larger.

{\it Conclusion.} The theoretical DR cross-section obtained in the previous study \cite{Ivan} was smaller by a factor 2-3 than the lowest measured experimental cross-section. However, the previous theory did not account for vibration along the CO coordinate, and the main purpose of the present study was to assess the validity of the frozen CO approximation employed there. It suggests that the CO vibration does not play a significant role in the DR process at energies below 0.1 eV, but starts to be important at higher energies, when the total energy of the ion+electron system approaches that of the first excited CO vibrational mode. This study suggests also that reduced dimensionality can be used in DR studies of small polyatomic ions as long as one includes (1) the dissociative coordinate and (2) the vibrational coordinates responsible for the highest probability of electron capturing. For HCO$^+$ the dominant dissociative coordinate is the CH bond (or $R_{G\rm{H}}$), whereas the vibrational coordinates responsible for the electron {\it capture} are $\theta$ and $\varphi$. Comparing with our previous study \cite{Ivan}, the effect of the CO vibration seems to be larger for DCO$^+$, which can be explained by a larger D/CO mass ratio.

In the present and previous theoretical studies, it was assumed that $s$- and $p$-wave-dominated eigenchannels are not mixed. This is justified to some extent by the fact that the {\it ab initio} energies used here account for the mixing, at least, at static geometries of the ion. However, in the dynamical framework of electron-ion collisions, the energy eigenstates obtained in the {\it ab initio} calculation could in principle be strongly mixed due to the permanent dipole moment ($\sim$3.5 debye) of HCO$^+$. The effect of the ionic dipole interaction with the electron should be addressed in future theoretical studies.

This work has been supported by the National Science Foundation under Grants No. PHY-0427460 and PHY-0427376, by an allocation of NERSC and NCSA (project \# PHY-040022) supercomputing resources.

\end{document}